\documentclass[conference]{IEEEtran}
\usepackage{graphicx,
	psfrag,
	epsfig,
	epstopdf,
    enumerate,
	amsthm,
    amsmath,
	amssymb,
	url,
	algorithm,
	algorithmic,
	balance,
	enumerate,
	color,
    url,
	tikz,
	pgfplots,
    geometry,
    placeins,
    cleveref,
    multirow,
    array,
}
\usepackage{amsmath}%[centertags][tbtags][sumlimits][nosumlimits][intlimits][namelimits][nonamelimits]
\usepackage[nospace,noadjust]{cite}
\newgeometry{left=1.6cm,right=1.6cm,bottom=1.6cm, top=1.6cm}

\newcommand{\ignore}[1]{}

\begin{document}
\title{Detecting Botnet Attacks in IoT Environments: An Optimized Machine Learning Approach}
\author{
\IEEEauthorblockN{MohammadNoor Injadat, Abdallah Moubayed, and Abdallah Shami}
		
\IEEEauthorblockA{Electrical and Computer Engineering Department, Western University, London, Ontario, Canada \\
	emails: \{amoubaye, minjadat, abdallah.shami\}@uwo.ca
}
%\IEEEauthorblockA{\IEEEauthorrefmark{2} Computer Engineering Dept., University of Sharjah, Sharjah, UAE \\
%	e-mail: anassif@sharjah.ac.ae
%
%}
}
\maketitle

\begin{abstract}
The increased reliance on the Internet and the corresponding surge in connectivity demand has led to a significant growth in Internet-of-Things (IoT) devices. The continued deployment of IoT devices has in turn led to an increase in network attacks due to the larger number of potential attack surfaces as illustrated by the recent reports that IoT malware attacks increased by 215.7\% from 10.3 million in 2017 to 32.7 million in 2018. This illustrates the increased vulnerability and susceptibility of IoT devices and networks. Therefore, there is a need for proper effective and efficient attack detection and mitigation techniques in such environments. Machine learning (ML) has emerged as one potential solution due to the abundance of data generated and available for IoT devices and networks. Hence, they have significant potential to be adopted for intrusion detection for IoT environments. To that end, this paper proposes an optimized ML-based framework consisting of a combination of Bayesian optimization Gaussian Process (BO-GP) algorithm and decision tree (DT) classification model to detect attacks on IoT devices in an effective and efficient manner. The performance of the proposed framework is evaluated using the Bot-IoT-2018 dataset. Experimental results show that the proposed optimized framework has a high detection accuracy, precision, recall, and F-score, highlighting its effectiveness and robustness for the detection of botnet attacks in IoT environments. 
\end{abstract}


\section{Introduction}\label{Intro_dns}
\indent The increased reliance on the Internet and the corresponding surge in connectivity demand has led to a significant growth in Internet-of-Things (IoT) devices. This is supported by the recent projections that the number of connected devices will reach around 28.5 billion devices by 2022 \cite{Cisco_data_growth}, with these IoT devices covering multiple use cases such as healthcare \cite{IoT_healthcare}, smart cities \cite{IoT_smart_cities}, and intelligent transportation systems \cite{IoT_ITS}. \\
\indent The continued deployment of IoT devices has in turn led to an increase in network attacks due to the larger number of potential attack surfaces. This is substantiated by Forbes' recent report stating that more than 2.9 billion events in 2019, a three-fold increase from the previous year \cite{IoT_attack_statistics}. Moreover, SonicWall reported that IoT malware attacks increased by 215.7\% from 10.3 million in 2017 to 32.7 million in 2018 \cite{IoT_attack_statistics1}. This illustrates the increased vulnerability and susceptibility of IoT devices and networks. Therefore, there is a crucial need for proper effective and efficient attack detection and mitigation techniques with researchers increasingly investigating and proposing multiple potential mechanisms \cite{SDP1,SDP2,IoT_attack_survey}.\\
\indent Machine learning (ML) has emerged as one potential solution due to the abundance of data generated and available for IoT devices and networks. ML allows systems to be dynamic and flexible to new inputs as they can “learn” without explicitly being told what to do \cite{Moubayed_EDM}. Additionally, ML techniques have illustrated their effectiveness and efficiency in various applications and use cases \cite{Moubayed_EDM1,Moubayed_EDM2,Injadat_EDM1,Injadat_EDM2,Moubayed_DNS1,Moubayed_DNS2}. Therefore, they have significant potential to be adopted for intrusion detection for IoT environments. Additionally, optimizing the parameters of these ML models is crucial to further enhance their detection performance and effectiveness \cite{Li_HPO,Moubayed_thesis,Injadat_thesis}.\\
\indent Therefore, this paper proposes an optimized ML-based framework consisting of a combination of Bayesian optimization Gaussian Process (BO-GP) and decision tree (DT) classification model to detect botnet attacks on IoT devices. The goal is to develop a dynamic, effective, and efficient IoT attack detection framework. As such, the main contributions of this work can be summarized as follows:
\begin{itemize}
	\item \textit{Proposing} a combination of BO-GP and DT classifier to detect botnet attacks on IoT devices.
	\item \textit{Evaluating} the performance of the proposed model using a recent  IoT dataset titled Bot-IoT-2018. 
\end{itemize}
%\begin{figure}[!h]
%	\centering
%	\includegraphics[scale=.5]{Figures/DNS_vulnerabilities_vertical.jpg}
%	%\includegraphics[trim=0.5cm 11.5cm 0.5cm 0.5cm, clip,scale=.35]{Figures/DNS_vulnerabilities.pdf}
%	\caption{DNS Vulnerabilities and Challenges}
%	\label{vulnerabilities}
%\end{figure}

\indent The remainder of this paper is organized as follows: Section \ref{related_work_dns} briefly surveys the literature. Section \ref{proposed_approach_dns} describes the proposed approach for IoT botnet detection and illustrates its complexity. Section \ref{dataset_description_dns} describes the dataset investigated. Section \ref{results_dns} presents the experimental setup and the obtained results. Finally, Section \ref{conc_dns} concludes the paper.
\section{Related Work}\label{related_work_dns}
\indent The recent surge in computing capabilities has led to an increase in investigating ML techniques and algorithms as an effective solution for network security  \cite{Li_IDS,Injadat_BO,Injadat_IDS1,Injadat_IDS2}. For example, Li \textit{et al.} proposed such models for intelligent transportation systems \cite{Li_IDS}. More specifically, the authors developed tree-based classification models in an attempt to detect intrusions in autonomous vehicles \cite{Li_IDS}. In contrast, Injadat \textit{et al.} proposed an optimized ML-based network intrusion detection framework using Bayesian optimization \cite{Injadat_BO}. Their experiments showed that the proposed model had a higher detection accuracy and a lower false alarm rate \cite{Injadat_BO}. In a similar manner, Injadt \textit{et al.} also proposed a multi-stage optimized ML-based intrusion detection framework that reduced the computational complexity while simultaneously improving the detection accuracy \cite{Injadat_IDS1}. Salo \textit{et al.} also proposed the use of ML classification techniques for intrusion detection \cite{Injadat_IDS2}. The authors proposed the combined use of ensemble feature selection and clustering-enabled classification models to detect unseen attack patterns \cite{Injadat_IDS2}.\\
\indent Within the specific context of IoT, multiple researchers proposed ML-based solutions to detection various attacks in such environments \cite{IoT_security_scada,IoT_security_DRNN,IoT_security_smart_homes}. Teixeira \textit{et al.} investigated five different ML classification algorithms to detect network attacks in an industrial IoT environment, namely in a  water storage tank’s control system \cite{IoT_security_scada}. In a similar fashion, Almiani \textit{et al.} proposed the use of deep recurrent neural networks to effectively detect network intrusions in IoT environments with the model showing high detection accuracy \cite{IoT_security_DRNN}. In contrast, Anthi \textit{et al.} proposed the use of various supervised ML classification models to detect four different types of IoT intrusion attacks in a smart home environment \cite{IoT_security_smart_homes}. The authors show that the proposed models had a high precision and recall values as well as a low classification time, illustrating the real-time deployment potential of these models \cite{IoT_security_smart_homes}. However, many of the frameworks dedicated to IoT environments use default parameters rather than optimized parameters for the different ML algorithms proposed as well as avoid addressing the class imbalance issue commonly found in network attack scenarios. As such, there is a need to further improve the detection performance by proposing optimized ML models.  
\section{Proposed Approach}\label{proposed_approach_dns}
\subsection{Proposed Approach Description}
\begin{figure}[!t]
	\centering
	\includegraphics[scale=.45]{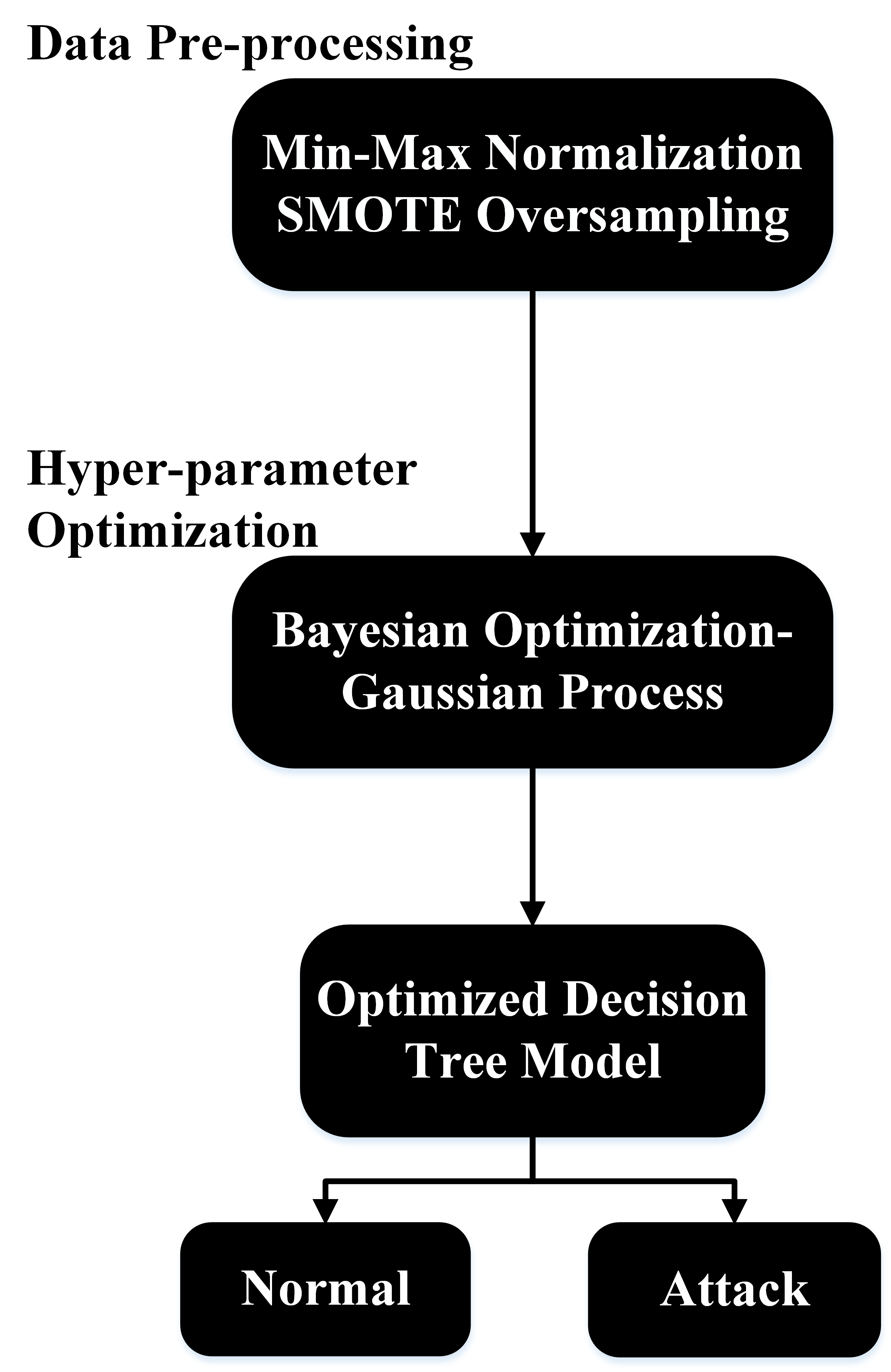}
	\caption{Proposed Optimized IoT Botnet Detection Framework}
	\label{iot_approach_fig}
\end{figure}

\indent This paper proposes combining the BO-GP algorithm to optimize the parameters of the DT classification model as part of an optimized ML-based framework for effective and efficient detection of botnet attacks on IoT devices. The proposed approach, as shown in Fig.\ref{iot_approach_fig}, consists of two components, namely:
\begin{enumerate}
	\item Data pre-processing: The goal of this component is to prepare the data into a format that would maximize the performance of the developed ML classification model. As such, this is done by initially normalizing the features using the min-max method. The feature normalization step helps unify the dynamic range of the different features so that no single feature dominates in the model training stage. This is done using the following equation \cite{Injadat_BO}:
	\begin{equation}
		x_{normalized}=\frac{x-min(x)}{max(x)-min(x)}
	\end{equation}

	After feature normalization, Synthetic Minority Oversampling TEchnique (SMOTE) is applied to oversample the minority class with the aim of addressing the class imbalance problem encountered in such environments \cite{imbalance1}. It is worth noting that SMOTE is proposed as it can  generate new high quality instances of the minority class \cite{SMOTE1}, resulting in an enhanced classification model performance and a reduced training sample size \cite{SMOTE1}.  
	\item Hyper-parameter Optimization: The goal of this component is to optimize the hyper-parameters of the ML model to ensure that the detection performance is maximized. To that end, BO-GP algorithm is proposed in this work to optimize the hyper-parameters of the DT model. BO-GP is one version of the Bayesian Optimization group of algorithms which belong to the class of probabilistic global optimization models \cite{Injadat_IDS1}. In our case, this is represented as the set of suitable values for the DT hyper-parameters. The BO-GP algorithm is chosen in this case since it can identify near-optimal hyper-parameter
	combinations within a few iterations \cite{Li_HPO}.
\end{enumerate}
Note that the proposed framework does not include a feature selection component. This is attributed to the fact that in many cases, data collected from IoT devices often contains a small number of bits and features due to their limited computing power. This is evident in multiple IoT deployment scenarios such as in smart homes \cite{IoT_security_smart_homes} and autonomous vehicles \cite{Li_IDS}.
\subsection{Complexity of Proposed Approach}
\indent The overall complexity of the proposed framework depends on the complexity of each of its components. Assume that the considered dataset is composed of $M$ instances and $N$ features. The complexity of the min-max feature normalization method is $O(N)$. This is because the algorithm determines the minimum and maximum of each feature to normalize the dataset. The complexity of the SMOTE algorithm is $O(M^2_{min}N)$ where $M_{min}$ is the number of minority class instances \cite{SMOTE_complexity}. The complexity of the BO-GP method for the hyper-parameter optimization stage is $O(M^3)$ \cite{Injadat_IDS1}. Finally, the complexity of the DT classification model is $O(M^2 N)$. However, the training time can be significantly reduced to approximately $O(\frac{M^2 N}{threads})$ where $threads$ is the maximum number of participating threads. This is because multi-threading is enabled when using this classifier \cite{Li_IDS}.  Therefore, the overall complexity of the proposed framework is $O(M^3)$, making it computationally efficient. 
%\indent The complexity of the proposed framework is dictated by the complexity of each stage. Assume that the dataset consists of $M$ samples and $N$ features. The complexity of the data pre-processing stage is governed by the complexity of the Z-score normalization method and the SMOTE method. The Z-score method has complexity of $O(N)$ given that it normalizes all the features using the respective means and standard deviation. In contrast, the SMOTE method has a complexity of $O(M^2_{min}N)$ where $M_{min}$ is the number of minority class instances \cite{SMOTE_complexity}. The complexity of the feature selection stage is that of the information gain method. This method has a complexity of $O(MN)$ since this method calculates the class-feature joint probabilities to determine the relevant features \cite{IGBFS_complexity}. Thirdly, the complexity of the PSO method for the hyper-parameter optimization stage is $O(N_{pop}N_{parm})$ where $N_{pop}$ is the size of the population assumed and $N_{parm}$ is the number of hyper-parameters to be optimized \cite{PSO_complexity}. Finally, the complexity of the optimized RF classification model is $O(\frac{M^2\sqrt{N_{red}}T}{Num_{thr}})$ where $N_{red}$ is the reduced feature set, $T$ is the number of trees, and $Num_{thr}$ is the number of threads on the computing device \cite{Li_IDS}. Therefore, the overall complexity of the proposed framework is $O(MN)$.
\section{Dataset Description}\label{dataset_description_dns}
\begin{figure}[!h]
	\centering
	\includegraphics[scale=.6]{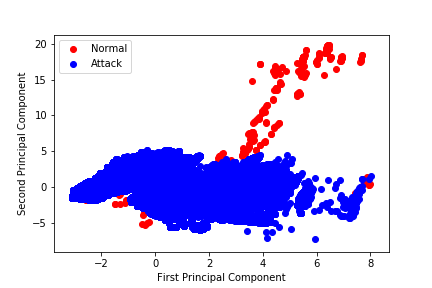}
	\caption{First and Second Principal Components of Bot-IoT-2018 Dataset}
	\label{pca_iot}
\end{figure}

\indent The dataset considered in this work is the Bot-IoT-2018 dataset developed in \cite{IoT_dataset}.  The dataset is built by designing a realistic IoT network environment using three main components:  network platforms, simulated IoT services, and feature extraction platform. The network platforms consist of different virtual machines (VMs) acting either as normal or attacking VMs. The simulated IoT services where generated using the Node-red tool to mimic the network behavior of IoT devices such as a weather station, a smart fridge, motion activated lights, garage door, and a smart thermostat. Finally, the feature extraction platform used Argus tool to extract the corresponding data features from the collected pcap files \cite{IoT_dataset}. The resulting dataset contains close to 72 million records and 46 features. For the purpose of this work, the reduced dataset consisting of around 3.6 million records (representing 5\% of the data as recommended by the dataset authors in \cite{IoT_dataset}) and the 10 best features is used for brevity. More specifically, the reduced dataset contains 477 \textbf{normal} instances and  3,668,045 \textbf{attack} instances. This illustrates that the dataset is significantly imbalanced. Fig. \ref{pca_iot} plots the first and second principal components of the dataset. It can be clearly seen that the dataset is highly non-linear and suffers from significant class imbalance, an issue commonly encountered in the network security field. This reiterates the need for applying data oversampling to ensure that the trained ML model has enough instances of each class to learn from.
\section{Experiment Results \& Discussion}\label{results_dns}
\subsection{Experiment Setup}
\indent The following settings were used to conduct the experiments in this work:
\begin{enumerate}
	\item Software: Python 3.7.4 running on Anaconda’s Jupyter Notebook.
	\item Hardware:  Intel\textsuperscript{\textregistered} Core\textsuperscript{TM} i7-9750H CPU 6 Cores at 2.6 GHZ and 16GB of memory running Windows 10.
\end{enumerate}
\subsection{Results \& Discussion}
\indent  To evaluate the performance of the proposed optimized DT-based framework, its performance is compared to a default DT classifier and the support vector machine (SVM) model proposed in \cite{IoT_dataset} using the following metrics \cite{Injadat_BO}:
\begin{equation}
	Accuracy=\frac{TP + TN}{TP+TN+FP+FN}
\end{equation}
\begin{equation}
	Precision=\frac{TP}{TP+FP}
\end{equation}
\begin{equation}
	Recall=\frac{TP}{TP+FN}
\end{equation}
\begin{equation}
	F-score=2*\frac{Precision*Recall}{Precision+Recall}
\end{equation}
where $TP$ is the number of true positive instances, $TN$ is the number of true negative instances, $FP$ is the number of false positive instances, and $FN$ is the number of false negative instances.\\
It can be seen in Table \ref{labeled_dataset_results} that the proposed optimized DT-based model outperforms the default model as well as the SVM model. More specifically, the proposed framework improved all the performance metrics including the testing accuracy, precision, recall, and F-score. For example, the results show that the default DT model achieves relatively low precision and high recall. This means that the model is incorrectly flagging normal traces as attacks. This is undesirable since this can lead to legitimate users being blocked or blacklisted. This behavior is attributed to the fact that the dataset used is significantly imbalanced with much more attack traces than normal ones. In a similar manner, the SVM model also had a lower recall value. In contrast, the proposed optimized DT-based framework was able to improve both the precision and recall values. This means that the model is effective in correctly identifying and differentiating between normal and attack instances. This is due to the oversampling performed to ensure that the model has enough instances of both classes to learn from. These results highlight the effectiveness and robustness of the proposed framework for botnet detection in IoT environments.   
\begin{table}[!t]
	\centering
	\caption{Performance Evaluation of Classifiers}
	\scalebox{0.85}{
		\begin{tabular}{|p{2.7cm}|p{1.4cm}|p{1.4cm}|p{1.4cm}|p{1.1cm}|}	\hline
			Algorithm & Testing $\;\;$ Accuracy (\%) & Precision & Recall&F-score\\ \hline
			Default DT&99.82&0.53&0.91&0.56\\ \hline
			SVM \cite{IoT_dataset}&88.37&1&0.88&0.94\\ \hline
			Optimized DT-based Framework&\textbf{99.99}&\textbf{0.99}&\textbf{1.00}&\textbf{1.00}\\ \hline
		\end{tabular}
	}
	\label{labeled_dataset_results}
\end{table}

\section{Conclusion \& Future Works}\label{conc_dns}
\indent A significant growth in the deployment of Internet-of-Things (IoT) devices has been observed in recent years due to the increased reliance on the Internet and the corresponding surge in connectivity demand. This is supported by the recent projections that the number of connected devices will reach around 28.5 billion devices by 2022. In turn, this has led to an increase in network attacks due to the larger number of potential attack surfaces. Therefore, proper effective and efficient attack detection and mitigation techniques are needed to ensure these devices are well protected.\\
\indent To that end, this paper proposed an optimized ML-based framework that combined Bayesian optimization Gaussian Process (BO-GP) and decision tree (DT) classification model to detect botnet attacks on IoT devices. The goal was to develop a dynamic, effective, and efficient IoT attack detection framework. Experimental results showed that the proposed optimized DT-based framework improved the accuracy, precision, recall, and F-score. More specifically, it achieved values of 99.99\%, 0.99, 1.00, and 1.00 for these four metrics respectively. This illustrated that the proposed framework is both effective and robust in detecting botnet attacks in IoT environments.\\
\indent This work can be extended in multiple directions. One intuitive direction is to use the complete dataset to ensure that more normal instances are used as part of the data oversampling process to further enrich the normal traces scenario. Another direction worth exploring is to investigate the time-related features to identify any temporal behaviors or patterns that may be helpful in detecting botnet attacks in IoT environments.

\small
\bibliographystyle{IEEEtran}
\bibliography{IoT_Ref}
\balance
\end{document}